\documentclass[natbib,vecphys]{svmult}

\usepackage{makeidx}         
\usepackage{graphicx}        
\usepackage{multicol}        
\usepackage[bottom]{footmisc}
\usepackage{amsmath}
\usepackage{url}
\usepackage{multirow}

\makeindex             

\begin{document}

\title*{Using Local Volume data to constrain Dark Matter dynamics}
\author{G. Lavaux\inst{1} \and R. Mohayaee\inst{1} \and S. Colombi\inst{1} \and R. B. Tully\inst{2}}
\institute{Universit\'e Paris 6/Institut d'Astrophysique de Paris,
  98bis Boulevard Arago, 75014 PARIS, FRANCE
\texttt{\url{lavaux@iap.fr}}
\and Institute for Astronomy, Univ. of Hawaii, Honolulu, USA }
%
%
\maketitle

\section{Introduction}

The peculiar velocities of galaxies hold a lot of information on the current
dynamical state of the mass tracers such as galaxies or groups of
galaxies in the Universe. This information can be used to constrain
the relationship between the mass tracers and the underlying dark
matter mass distribution responsible of the observed dynamic. This
gives a measurement of $\beta$, the linear growth factor
of density fluctuations, which yields either a measure of
$\Omega_\text{m}$, the mean matter density in the Universe, or of the
shape of the function $M/L (L)$. We
propose here to show that Lagrangian reconstructions of peculiar velocities permit this kind
of comparison and may serve to constrain the above two quantities. However these constraints can only be
tight if observational biases are taken into account correctly, which is in most cases possible.
Galaxies of the Local Volume, with good peculiar velocity
measurements, may thus help constraining the dark matter dynamics.

Lagrangian reconstruction predicts peculiar velocities of individual
mass tracers from only their redshift
position and their mass. We will illustrate the
reconstruction procedure by the means of the
Monge-Amp\`ere-Kantorovitch (MAK) reconstruction, which has already
been extensively tested on simulations \citep{Brenier2002, moh2005}.
The impact of observational biases such as redshift distortion, mass-to-light
assignment (diffuse mass, $M/L$ relation), incompleteness effect,
finite volume effects, statistical measurement of $\beta$
has only been studied thoroughly in \cite{ObsBias07}. 
In this proceeding, we propose to look more precisely at the consequence of
choosing a specific mass-to-light assignment on detected tracers. 

To study the aforementioned effects, we use a 128$^3$ N-body
collisionless sample in a 200$^3$~$h^{-3}$Mpc$^3$ volume \citep{moh2005}, with $\Omega_\text{m}=0.30$, $\Omega_\Lambda=0.70$. 
From a halo catalogue, {\it FullMock}, built from the sample, we have produced mock catalogues.
{\it FullMock} is
obtained using a standard Friend-of-friend algorithm, whose linking
parameter is chosen to be a fifth of the mean particle separation of
the original sample. Haloes of less than 5 particles were left unbound
and the corresponding particles put in a set called the ``background
field''. This set represents 63\% of the total mass of the
simulation.

\section{Correctable observational biases}
\label{sec:correctable}

\begin{figure}
  \begin{center}
    \newsavebox{\onefigure}
    \newlength{\myfigh}
    \sbox{\onefigure}{\includegraphics[width=.3\hsize]{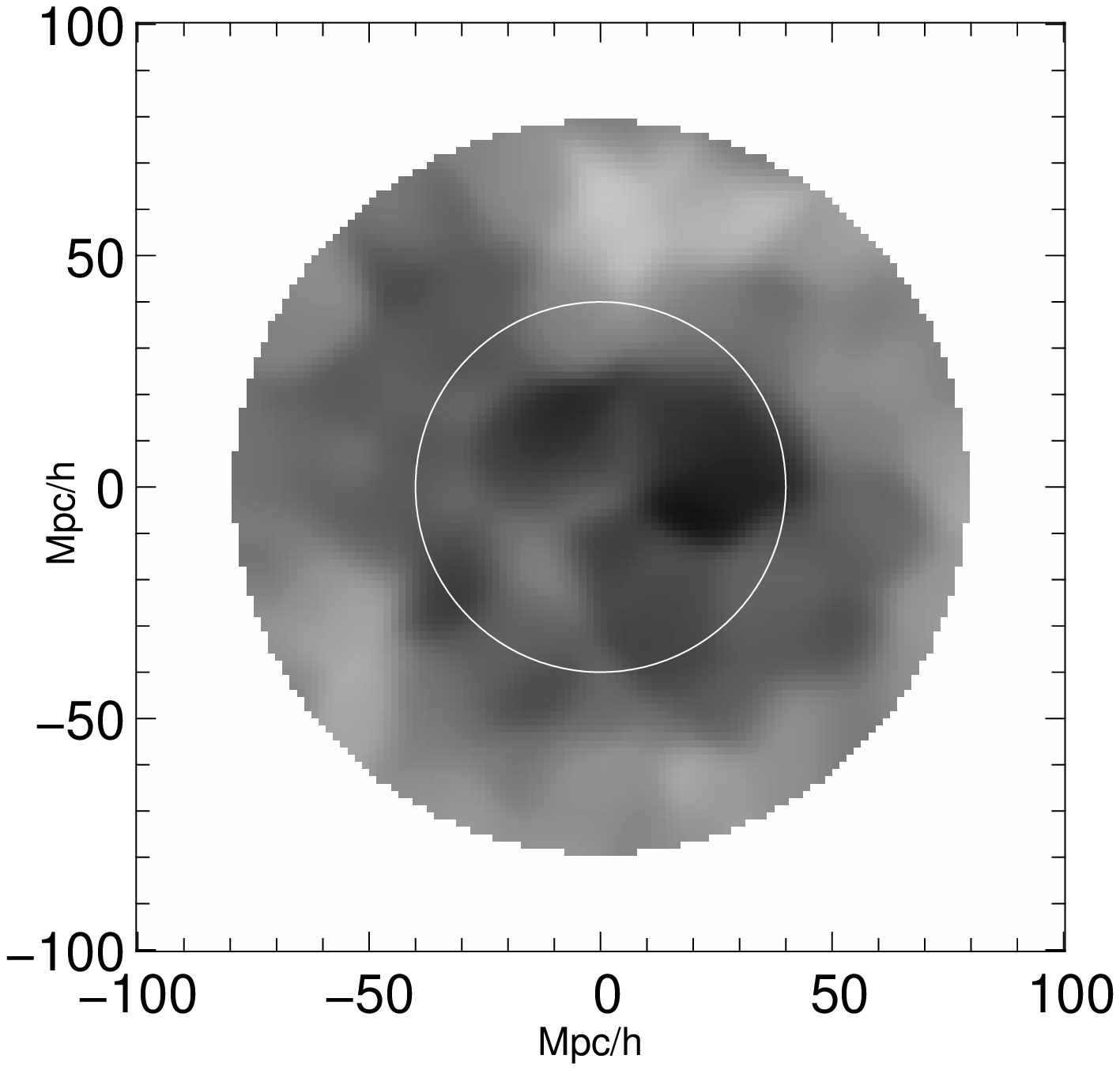}}
    \settoheight{\myfigh}{\usebox{\onefigure}}
    \begin{tabular}{cccc}
      &
      \parbox{.3\hsize}{\centering Simulated \\ velocity field} & 
      \parbox{.3\hsize}{\centering Reconstructed \\ velocity field} &
      \parbox{.3\hsize}{\centering Individual object \\ comparison} \\
      \includegraphics[height=\myfigh]{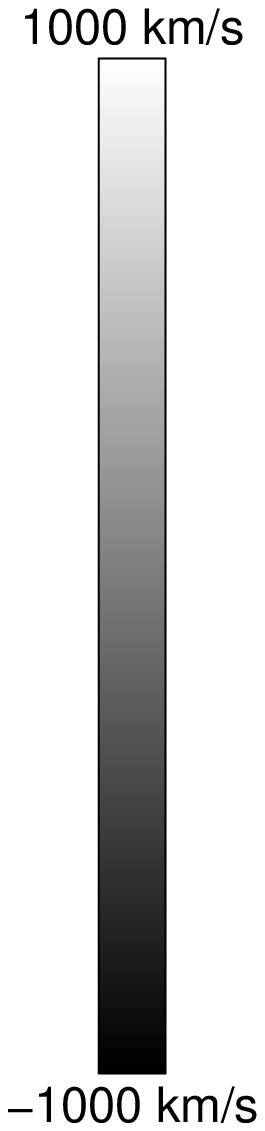} &  \includegraphics[width=.3\hsize]{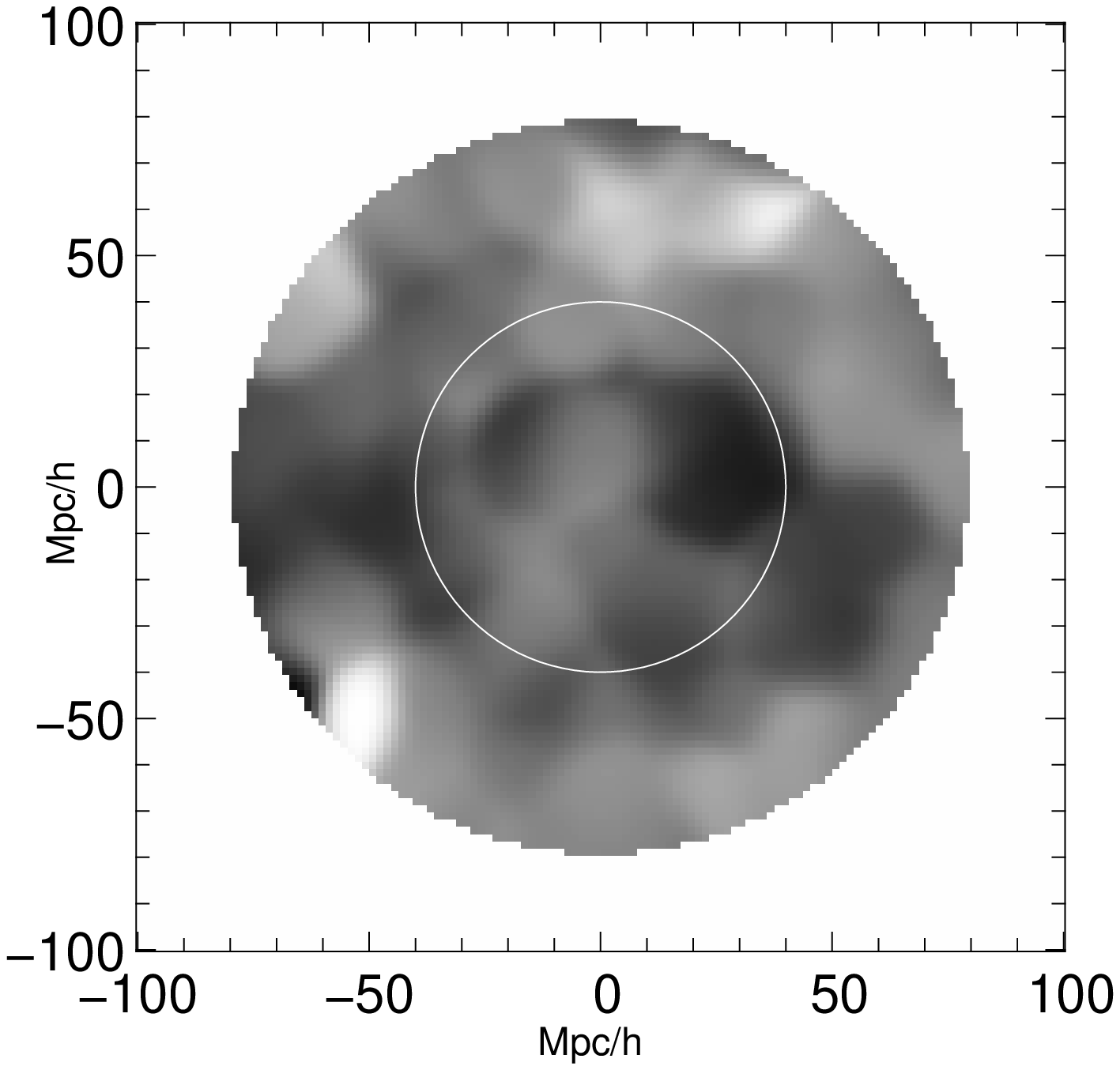} &
      \includegraphics[width=.3\hsize]{rec_field} &
      \includegraphics[width=.3\hsize]{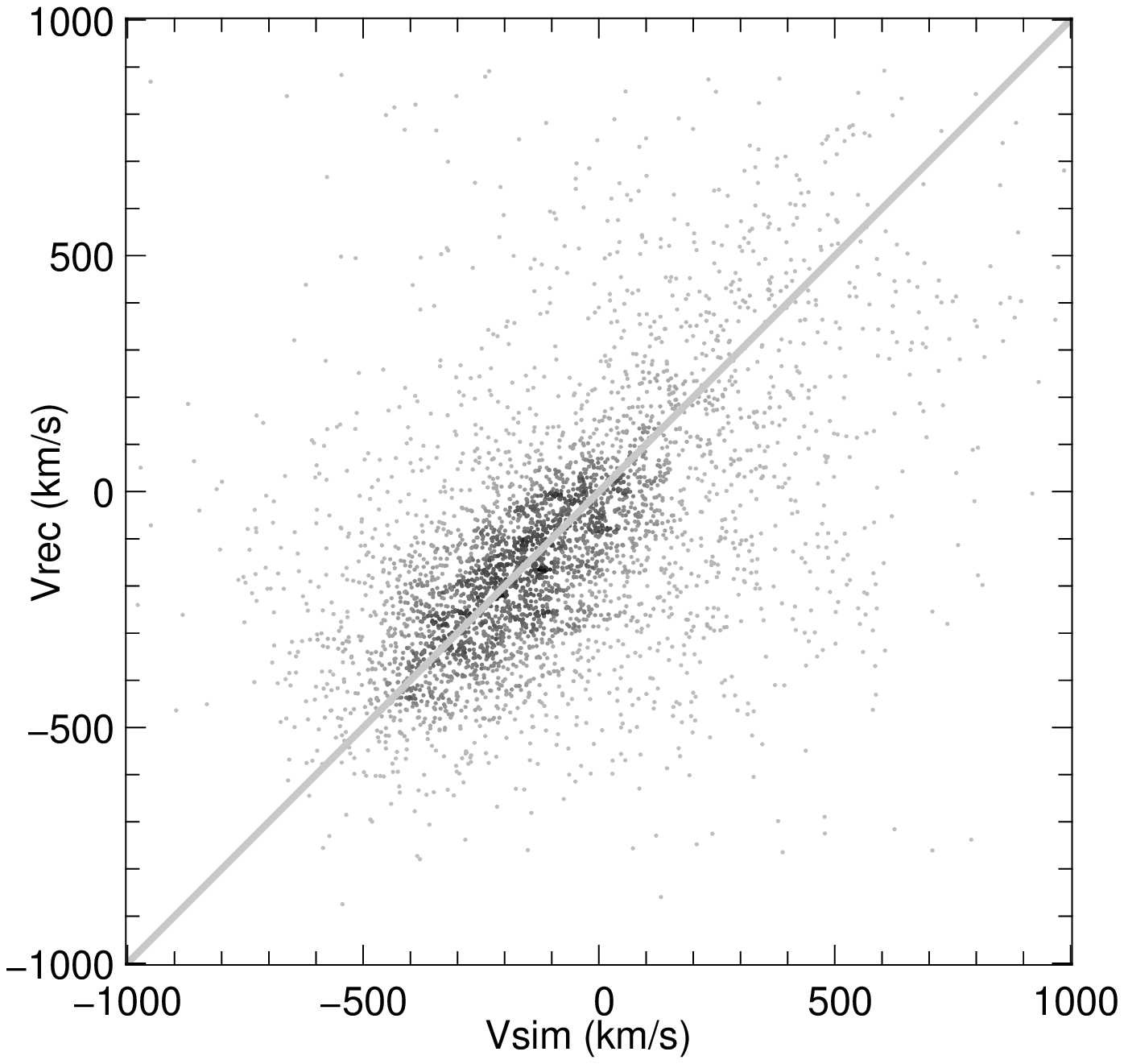}
    \end{tabular}
  \end{center}
  \caption{\label{fig:everything} Result of a reconstruction including finite volume effects (selection and boundary effects), redshift distortion, incompleteness. Left panel: simulated velocity field
  smoothed with a 5~$h^{-1}$Mpc Gaussian filter. The white circle shows the
  40~$h^{-1}$Mpc region inside the 80~$h^{-1}$Mpc deep mock catalogue. Middle panel: reconstructed
  velocity field smoothed equally. Right panel: individual comparison between reconstructed and simulated velocities for haloes of the mock catalogue inside the white circle.\vspace{-.25cm}}
\end{figure}

The result of including all the observational effects mentioned in
the introduction, except the mass-to-light assignment problem, in
mock catalogues are given in Fig.~\ref{fig:everything}. As one can see
for a standard mock catalogue, built to have approximately the same
features as observed redshift catalogues (such as 2MASS Redshift Survey, 2MRS hereafter), there is
no significant bias introduced on reconstructed velocities. However the
scatter may be significant, all the more that the number of
measurements is low, which means a good statistical description of the
error is needed to achieve an unbiased Bayesian estimator of
$\Omega_\text{m}$ from measurements. A problem that is not highlighted
in this figure, is that incompleteness correction, if done unwisely,
{\it e.g.} by assuming a wrong $\Omega_\text{m}$,
may yield an offset in the relation shown in the right
panel of Fig.~\ref{fig:everything}. Unfortunately, a similar offset is
introduced by a wrong choice of the Hubble constant $H$ and thus its measurement may be hindered by incompleteness effects though hopefully without strongly affecting the $\Omega_\text{m}$ measurement.

\section{Mass-to-light assignment}
\label{sec:ml}

\subparagraph{Diffuse mass} -- The first problem comes from the intrinsic limitation of redshift catalogues:
they have a lower luminosity cut-off which translates to a minimum
lower mass. In fact, in 2MRS, with $L_\text{min,K} \simeq 10^9 \text{ L}_\odot$, and
using \cite{ShethTormen02}, we find that $\simeq 50\%$ of the mass distribution may be missing in real data. We have
tried to reproduce that problem in {\it FullMock} by not using
the background field to build the reconstruction mesh. 
It has lead
to an overestimation of the reconstructed peculiar velocities because
the gravitational field of the haloes is in reality screened by the
background field.
On the contrary, if one predicts the amount of
missing mass and introduces it
homogeneously in the reconstruction mesh, reconstructed peculiar velocities
are underestimated because of a too important screening effect. This two extreme cases lead, for mock
catalogues, to low constraints on $\Omega_\text{m}$: $0.15 \le
\Omega_\text{m} \le 0.74$. However we showed there is an
optimal compromise between those two situations that give an unbiased
reconstruction, with nearly no extra scatter on reconstructed
velocities, when compared to a reconstruction on {\it FullMock}.

\subparagraph{$M/L$ relation} -- Another fundamental problem of mass-to-light assignment is the use
of a unique $M/L$ relation to transform detected galactic
luminosities to dynamical masses. We have built a luminosity catalogue
from {\it FullMock} using the $M/L$ relation given by
\cite{TullyML05}. A mass tracer catalogue is then recovered by
assuming either a constant $M/L$ or a $M/L$ as given by
\cite{Marinoni02} which is obtained by mapping a Press-Schechter mass function to a
Schechter luminosity function. The reconstructed velocities are then
compared to the simulated ones. We tried also to introduce a
significant uniform scatter on the logarithm of the mass $M$ of haloes in
the original mock catalogue. The width of the distribution was chosen with
$\Delta \log_{10} M = 1$ to be consistent with observations. The
result of these three tests is that a random uncertainty on the $M/L$
relation does not produce bias and leads only to a relatively small
increase of the scatter between reconstructed and simulated
velocities, whereas changing the relation (even a little) quickly introduces
a significant systematic when comparing these same velocities.

\section{Application to an Extended Nearby Galaxies Catalog}

\begin{figure}
  \begin{center}
    \begin{tabular}{ccc}
      Reconstructed velocities & Measured velocities & Within 20~$h^{-1}$Mpc \\
      \includegraphics*[width=.3\hsize]{3kvelmap_rec} &
      \includegraphics*[width=.3\hsize]{3kvelmap_mes} &
      \includegraphics[width=.285\hsize]{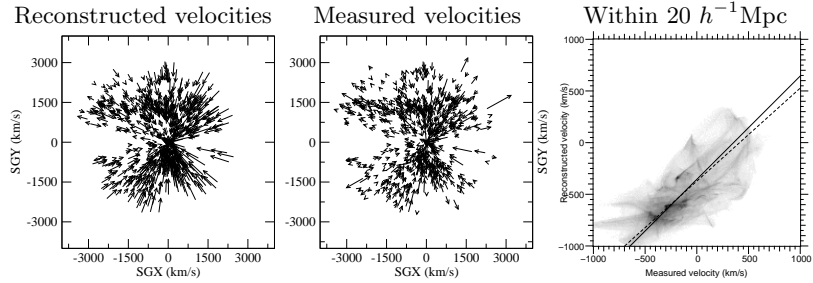}
    \end{tabular}
  \end{center}
  \vspace{-.15cm}
  \caption{\label{fig:3k} Left panel: line-of-sight component of
    reconstructed velocities for objects with a measured
    velocity. Middle panel: same as left panel but for the measured
    velocities. Righ panel: Comparison between reconstructed and measured
    velocity field, obtained after adaptive smoothing of
    the corresponding individual velocities on a uniform grid. Solid line corresponds to $\Omega_\text{m}=0.30$, dashed-line to $\Omega_\text{m}=0.38$ (result given by the Bayesian estimator).}\vspace{-.25cm}
\end{figure}

While these results are still preliminary (Lavaux et al. in
preparation), we present here a MAK reconstruction for which we have tried
to account for redshift distortion, zone of avoidance, finite volume
effect and incompleteness. We chose $M/L=300$ for elliptical galaxies
and $M/L=100$ for spiral galaxies. This choice is motivated by
observations of velocity dispersion of groups.
As we have neglected the influence of
the diffuse mass, we are most likely going to overestimate
reconstructed peculiar velocities and thus to underestimate
$\Omega_\text{m}$, as shown in \S~\ref{sec:ml}.
Fig.~\ref{fig:3k} summarises the result obtained using this
reconstruction. The reconstructed line-of-sight peculiar velocities
look in agreement with measurements, at least in the central
region. Of course, in the outer part of the catalogue measured velocities are
both intrinsically noisy and incorrectly reconstructed because of
boundary effects. The right panel shows that reconstructed velocities
are well correlated to measurements but that there is an
offset most probably coming from an incompleteness correction defect, mentioned in \S~\ref{sec:correctable},
that needs be corrected in future reconstructions. 

\section{Conclusion and perspective}

So far we have checked the influence of each observational bias and
shown what must be included to extract useful informations
on dark matter dynamics. The first applications of our method to real
galaxy catalogues look successful though are still probably affected by
some of the biases. More checking must be conducted as has been
done on mock catalogues to establish the amount of systematics. We are
now going to apply the method to a bigger catalogue like 2MRS or 6dF
to decrease finite volume effects. Other information may be
extracted from Lagrangian reconstruction of the Local Volume such as
local initial conditions for constrained simulations and
statistical quantities (power 
spectrum, cosmological parameters) that may be compared to those obtained by CMB
experiments. 

%
%

\begin{thebibliography}{6}

\bibitem[{Brenier} et~al.(2003){Brenier}, {Frisch}, {H{\'e}non}, {Loeper},
  {Matarrese}, {Mohayaee}, and {Sobolevski{\u i}}]{Brenier2002}
Y.~{Brenier}, U.~{Frisch}, M.~{H{\'e}non}, G.~{Loeper}, S.~{Matarrese},
  R.~{Mohayaee}, and A.~{Sobolevski{\u i}}.
\newblock \emph{MNRAS}, \textbf{346}, 501 (2003).

\bibitem[{Mohayaee} et~al.(2006){Mohayaee}, {Mathis}, {Colombi}, and
  {Silk}]{moh2005}
R.~{Mohayaee}, H.~{Mathis}, S.~{Colombi}, and J.~{Silk}.
\newblock \emph{MNRAS}, \textbf{365}, 939 (2006).

\bibitem[{Lavaux} et~al.(2007){Lavaux}, {Mohayaee}, {Colombi}, {Tully},
  {Bernardeau}, and {Silk}]{ObsBias07}
G.~{Lavaux}, R.~{Mohayaee}, S.~{Colombi}, R.~B. {Tully}, F.~{Bernardeau}, and
  J.~{Silk}.
\newblock \emph{ArXiv e-prints}, 707 (2007).

\bibitem[{Sheth} and {Tormen}(2002)]{ShethTormen02}
R.~K. {Sheth} and G.~{Tormen}.
\newblock \emph{MNRAS}, \textbf{329}, 61 (2002).

\bibitem[{Tully}(2005)]{TullyML05}
R.~B. {Tully}.
\newblock \emph{ApJ}, \textbf{618}, 214 (2005).

\bibitem[{Marinoni} and {Hudson}(2002)]{Marinoni02}
C.~{Marinoni} and M.~J. {Hudson}.
\newblock \emph{ApJ}, \textbf{569}, 101 (2002).

\end{thebibliography}

%


\printindex
\end{document}